\documentclass[twocolumn,amsmath,amssymb,aps,prstab,floatfix]{revtex4}
\usepackage[latin9]{inputenc}
\usepackage{color}
\usepackage{amstext}
\usepackage{amssymb}
\usepackage{graphicx}
\usepackage{soul}
\usepackage{epstopdf}
\usepackage[version=3]{mhchem}
\newcommand*{\rom}[1]{\expandafter\@slowromancap\romannumeral #1@}
\usepackage{pgfplots}

\makeatletter
\@ifundefined{textcolor}{}
{%
 \definecolor{BLACK}{gray}{0}
 \definecolor{WHITE}{gray}{1}
 \definecolor{RED}{rgb}{1,0,0}
 \definecolor{GREEN}{rgb}{0,1,0}
 \definecolor{BLUE}{rgb}{0,0,1}
 \definecolor{CYAN}{cmyk}{1,0,0,0}
 \definecolor{MAGENTA}{cmyk}{0,1,0,0}
 \definecolor{YELLOW}{cmyk}{0,0,1,0}
}

\makeatother

\begin{document}

\title{Dynamics of Micro-vortices Induced by Ion Concentration Polarization}

\author{Joeri de Valenca$^{~a,b~\S}$}

\author{R.M. Wagterveld$^{~b}$}

\author{Rob G.H. Lammertink$^{~a}$}

\author{Peichun Amy Tsai$^{~a, c~\ast}$}

\affiliation{{$^{a}$Soft Matter, Fluidics and Interfaces Group, MESA$^{+}$
Institute, University of Twente, 7500~AE Enschede, The Netherlands;
$^{b}$Wetsus, European Centre of Excellence for Sustainable Water Technology,
Oostergoweg 9, 8911 MA Leeuwarden, The Netherlands; $^{c}$Department of Mechanical Engineering, University of Alberta, Edmonton, Alberta T6G 2G8, Canada}}

\date{\today}
\begin{abstract}
We investigate the coupled dynamics of the local hydrodynamics and global electric response of an electrodialysis system, which consists of an electrolyte solution adjacent to a charge selective membrane under electric forcing. Under a DC electric current, counterions transport through the charged membrane while the passage of co-ions is restricted, thereby developing ion concentration polarization (ICP) or gradients. At sufficiently large currents, simultaneous measurements of voltage drop and flow field reveal several distinct dynamic regimes. Initially, the electrodialysis system displays a steady Ohmic voltage difference ($\Delta V_{ohm}$), followed by a constant voltage jump ($\Delta V_c$). Immediately after this voltage increase, micro-vortices set in and grow both in size and speed with time. After this growth, the resultant voltage levels off around a fixed value. The average vortex size and speed stabilize as well, while the individual vortices become unsteady and dynamic. These quantitative results reveal that micro-vortices set in with an excess voltage drop (above $\Delta V_{ohm} + \Delta V_c$) and sustain an approximately constant electrical conductivity, destroying the initial ICP with significantly low viscous dissipation. 
\end{abstract}
\maketitle

\vspace{-0.2in}
Ion concentration gradients emerge during a separation process involving a charge selective surface (electrode or membrane) in an electrolyte solution, inducing a decreasing ion concentration towards the interface and thereby hampering ion transport. For example, for water purification using electrodialysis under electric forcing, the charge selectivity of an ion exchange membrane causes an enrichment of counter-ions on the permeated side of membrane and a depletion of co-ions on the feed side. This so-called ion concentration polarization (ICP), with a decreasing ion concentration on the (feed side) interface, is a common theme in electrochemical applications that influence the performance of ion separation and transport. The diversity of ICP-associated applications has recently motivated numerous studies, ranging from micro-and-nano-junctions~\cite{Kim_2007,Yossifon_2008, Han_2010, Kim_2012,Chang_2012,Jarrod_2014, Nielsen2014, Green_2015}, electrodialysis~\cite{Rubinstein_2008, Nikonenko_2010, Kwak_2013, Wessling_2014, abu-Rjal_2014}, desalination~\cite{Mani_2011, Nikonenko_2014}, electrodeposition~\cite{Fleury1991,Huth1995,Rosso2007,Gonzalez2007}, and fuel cells~\cite{Scott_1999}.  

\begin{figure}[h]
\includegraphics[width=0.95\columnwidth]{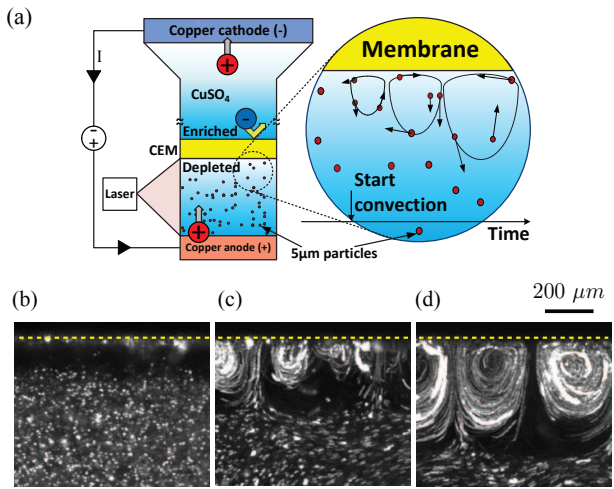}
\protect\protect\caption{(a) Schematic diagram of the experimental setup of a charge selective (cation exchange) membrane, which allows cations to pass the membrane while rejecting anions, immersed in an ion solution of \ce{CuSO4} under an electric DC current between two copper electrodes. We simultaneously measure the global  voltage drop, $\Delta V$, under a constant applied current, $I$, and local hydrodynamics of the ionic solution. (b--d) Representative motions of the seeding micro-particles reveal different hydrodynamic states in time: (b) Electrical conduction regime without hydrodynamic convection (at 100~s). This transforms to a convection regime with micro-vortices growing in size with time, (c--d) 220~s and 420~s, respectively. The applied DC current density, $J$, was 10~A/m$^{2}$. The dashed lines indicate the bottom boundary of the membrane.}
 \label{fig1} 
\end{figure}

In electrodialysis, theoretical analyses of low dimensions reveal that ICP can drive hydrodynamic instability through an equilibrium~\cite{Rubinstein_2015} or non-equilibrium (electro-osmotic or bulk electro-convective) mechanism~\cite{Dukhin_1991,Rubinstein_1988,Rubinstein_2000,Leinweber_2004,Rubinstein_2005,Rubinstein_2008,Rubinstein_2010}, suggesting an additional charge transport due to ICP-induced fluid motion under a sufficiently large DC voltage. Recent advances have been made with direct numerical simulations (DNS)~\cite{Pham_2012, Druzgalski_2013,Demekhin_2013}, providing insights into ion concentrations and flow velocity adjacent to a charged membrane. Experimentally, under a pressure-driven micro-channel flow, the advection and height selection of the unidirectional sheared vortices were characterized along the membrane~\cite{Kwak_2013}, but the internal vortex structure was not probed. Furthermore, quantitative experiments of the flow field in electrodialysis without shear flow are still missing, in particular under high electrical forcing~\cite{Rubinstein_2008}. In this paper, we show the first quantitative measurements of the coupled hydrodynamics and electrical response of an ionic solution in the vicinity of a charge selective membrane, under a constant DC electrical current without an external shear flow.

Fig.~\ref{fig1} shows the experimental setup and the resulting fluid dynamics at high electric forcing in electrodialysis.  We used a cation exchange membrane, CEM (Neosepta CMX, surface area of 3~mm $\times$ 4~mm with thickness of $170~\mu$m), horizontally placed in a transparent PMMA cell filled with a $10$mM~\ce{CuSO4} electrolyte between two copper electrodes. We performed chronopotentiometric measurements between the top cathode and the bottom anode with an electrometer (Autolab PGSTAT30 Potentiostat). This method consists of forcing a constant DC electric current, $I$, through the ionic solution across the membrane and measuring the time-dependent voltage difference, $\Delta V(t)$ between the electrodes. The bottom anode, where copper oxidizes, serves as a cation source; the top cathode, where copper reduces, acts as a cation sink.  This configuration of the electrodes can suppress the occurrence of gravitational convection caused by a variation of fluid density due to ion-concentration since a heavier \ce{Cu^{2+}}-rich solution is present close to the anode at the bottom of the fluid cell~\cite{Hage_2010}. We noticed that Cu dendrites form on the cathode after long experimental runs ($\ge$ 1000 s) and at a high current density. Thus, electrodes are cleaned before each experiment. We focused on the time-series data before the maximal ICP condition for the cathode.

In addition to the electrical measurements $\Delta V(t)$, simultaneously, the flow motion is observed close to the membrane interface under different DC currents (see Fig.~\ref{fig1}b--d).  The flow motions are observed by seeding $0.1$~wt$\%$ $5~\mu$m polystyrene tracer particles (Microparticles GmbH, with the particle density of $1.05$~g/cm$^{3}$) to the solution.  For these nearly buoyancy-neutral micro-particles, the theoretical sedimentation speed is small ($\approx0.7\,\mu m/s$, estimated by the Stokes drag equation~\cite{Raffel_2007}). The zeta potential of these micro-particles is measured to be $\approx-1$~mV for 1~mM~\ce{CuSO4}~(Zetasizer Nano ZS, Malvern). These micro-particles do not influence the electrical response of the electrolyte solution, as shown by similar $\Delta V(t)$ data obtained with and without the micro-particles. To avoid particle aggregation, a non-ionic surfactant, Tween80, ($0.1\, wt\%$) is added to the solution.  We obtain accurate flow fields employing a particle image velocimetry (PIV) technique~\cite{Raffel_2007, Tsai_PoF_2009}. The micro-particles are illuminated by a thin laser sheet (Firefly laser, $808~n$m, Oxford Lasers) with a pulse duration of $20~\mu s$ and a pulse power of $0.3~$mJ/pulse (illuminating 3~mm $\times$ 2~mm $\times$ 200~$\mu$m). The scattered light is captured at 20~Hz by a CCD camera (Sony XCG-H280E, 1920$\times1080$ px$^2$), with a magnifying lens (Navitar, 2--14 $\times$) placed perpendicular to the laser sheet. Sets of 50-200 images are analyzed using ImageJ software (NIH) to visualize and measure the vortex motions and sizes.

We also determine the vortex speed and size using PIV analysis, with a typical time delay (of 0.1 sec) between the image pair.  The focal depth of the optical system is measured to be $\approx 200{\ensuremath{\mu}m}$. Particles outside this depth of field are larger, have a lower light intensity, and are systematically filtered in the PIV analysis. To calculate the flow field, we use a multi-grid cross correlation method with decreasing window size by $\approx 50\%$~\cite{Karatay_PNAS_2013, LaVision}. First we use a 128 $\times$ 128 px$^2$ interrogation window to determine a reference vector field. This field is subsequently used to calculate a (window) shift for the next correlation. 
To get a higher resolution, the second calculation is done with windows of 96 $\times$ 96 px$^2$, and the vectors are displayed with a 50\% overlap~\cite{Karatay_PNAS_2013}. From these vector fields, we determine the mixing layer thickness, $L_{mix}$, of the vortex region and the root mean square velocity (average vortex speed) within this layer.

Each experiment starts with a uniform concentration of ~\ce{CuSO4} at both sides of the membrane. As a DC electric current is forced through the charged membrane, the counterions (\ce{Cu^{2+}} cations), which can easily pass through the membrane, migrate upwards, whereas co-ions (\ce{SO4^2-} anions) migrate downwards. Because the co-ions are retained by the charged membrane, the co-ion concentration enriches at the upper (cathode) side and depletes at the bottom (anode) side of the membrane. At a critical condition, a maximum ion-depletion occurs, with a vanishing co-ion concentration at the membrane surface. This so-called limiting current density can be estimated by balancing electromigration with diffusion of the co-ions~\cite{Krol_1999, Nikonenko_2010}, and $J_{lim} = 3.3~A$/m$^2$ for our electrodialysis system. We apply a current density, $J$, above this limit to study the `over-limiting' conductivity induced by ICP, a long-standing unsolved problem.


\begin{figure}
\includegraphics[width=0.9\columnwidth]{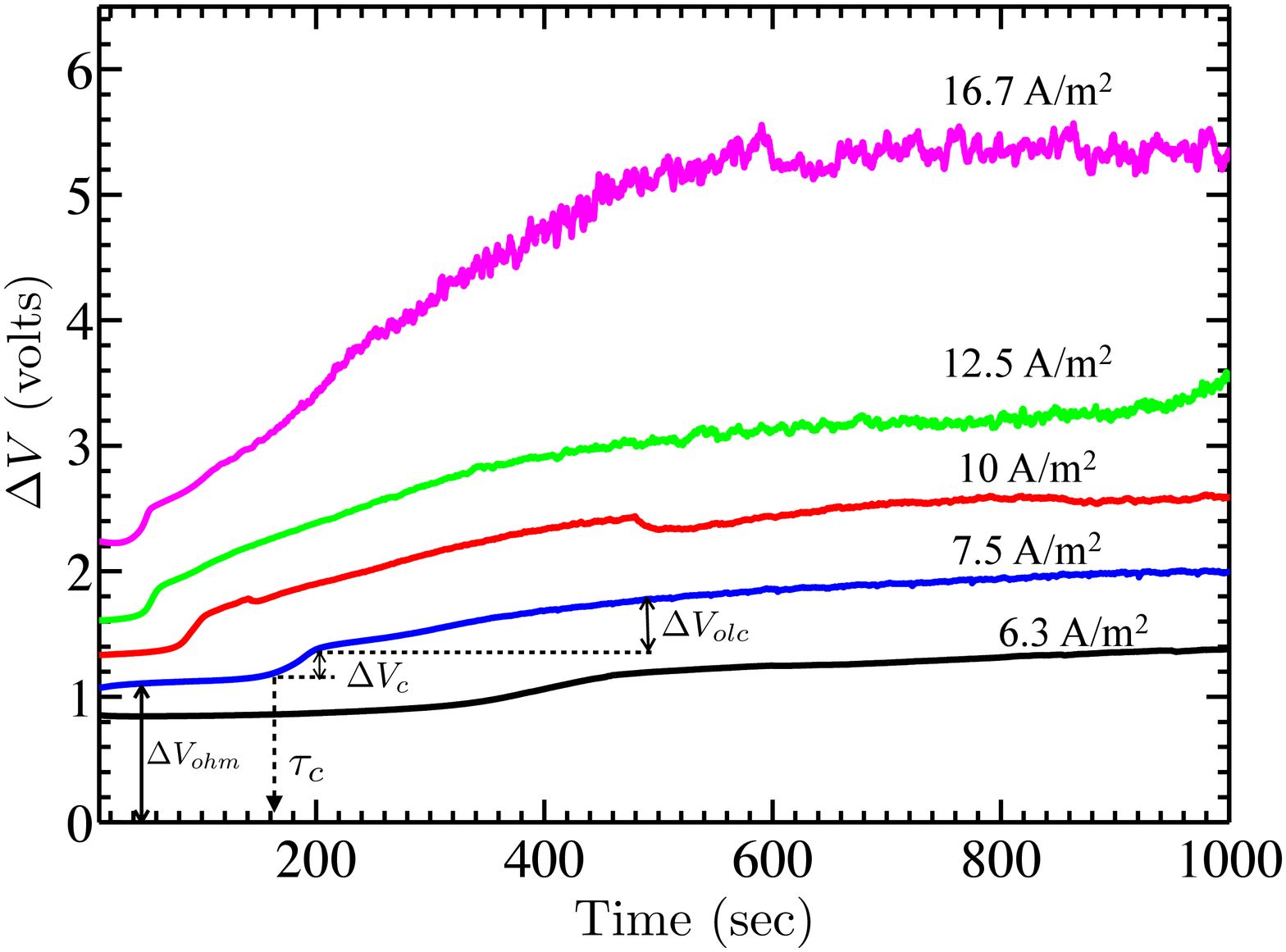} 
\protect\protect\caption{Chronopotentiometric measurements of the total electric voltage drop,  $\Delta V(t)$ changing in time across the cell, under different applied DC current densities, $J$, through the charge selective membrane. The total potential drop  reveals three distinct responses: (1) The initial constant Ohmic resistance of the whole cell $\Delta V_{ohm}=IR_{ohm}$, (2) The critical voltage jump, $\Delta V_{c}$ that starts at the transitional time, $\tau_{c}$, and (3) The over-limiting voltage drop over the vortex mixing region, $\Delta V_{olc}$.}
\label{fig2} 
\end{figure}

Fig.~\ref{fig2} shows the experimental results of the global electrical responses of the electrodialysis system. 
In time, the redox reactions at the electrodes start immediately, and ion concentration polarization is gradually developed.  The initial constant voltages ($\Delta V_{ohm}$) reflects a constant Ohmic resistance of the electrodialysis system: $R=11.2\, k\Omega$, obtained from a linear fit of $\Delta V_{ohm}$ for different $I$. Subsequently, ion concentrations are depleted close to the anode side of the membrane.  The depletion grows until the critical limiting condition, which leads to an increased electrical resistance. This critical condition is manifested by a sharp increase in measured voltage, $\Delta V_{c}$, after a transitional time $\tau_{c}$.  By the same token for $J_{lim}$ described above, $\tau_{c}$ can be estimated using Fick's second law with a vanishing co-ion concentration ($c_{-}$) at the membrane surface, i.e., Sand's equation~\cite{Krol_1999, Nikonenko_2010}. In agreement with the theory, our $\tau_{c}$ has a linear relationship with $1/I^{2}$, with a fitted cation transport number in the membrane of 0.9, consistent with the previous experimental results~\cite{Dugoecki_2010, Krol_1999, Nikonenko_2010}. The voltage jump $\Delta V_{c}$ corresponds to the (electrical current) plateau region in the $IV$ curve under a DC voltage (e.g. Fig.~4 in Ref.~\cite{Krol_1999}) and depends on the type of the membrane. We measured $\Delta V_{c}=0.3\pm0.03~V$ for different $J$, ranging from 6.3 to 16.7~A/m$^{2}$. The jump is followed by a further voltage increase, $\Delta V_{olc}$, until $\Delta V$ reaches a saturated value at a later time. 

\begin{figure}[ht]
\begin{centering}
\includegraphics[width=3.0in]{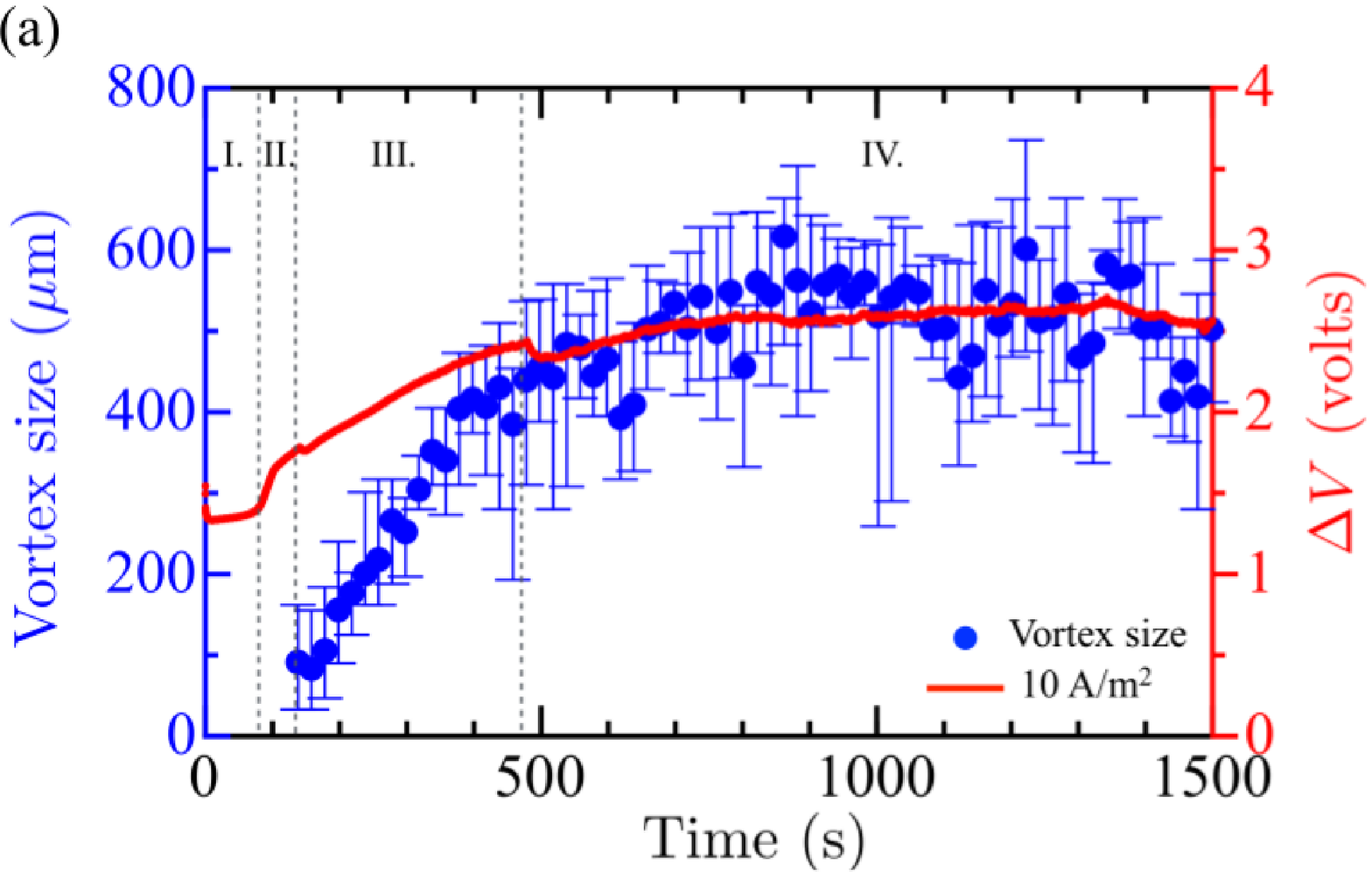} 
\includegraphics[width=2.4in]{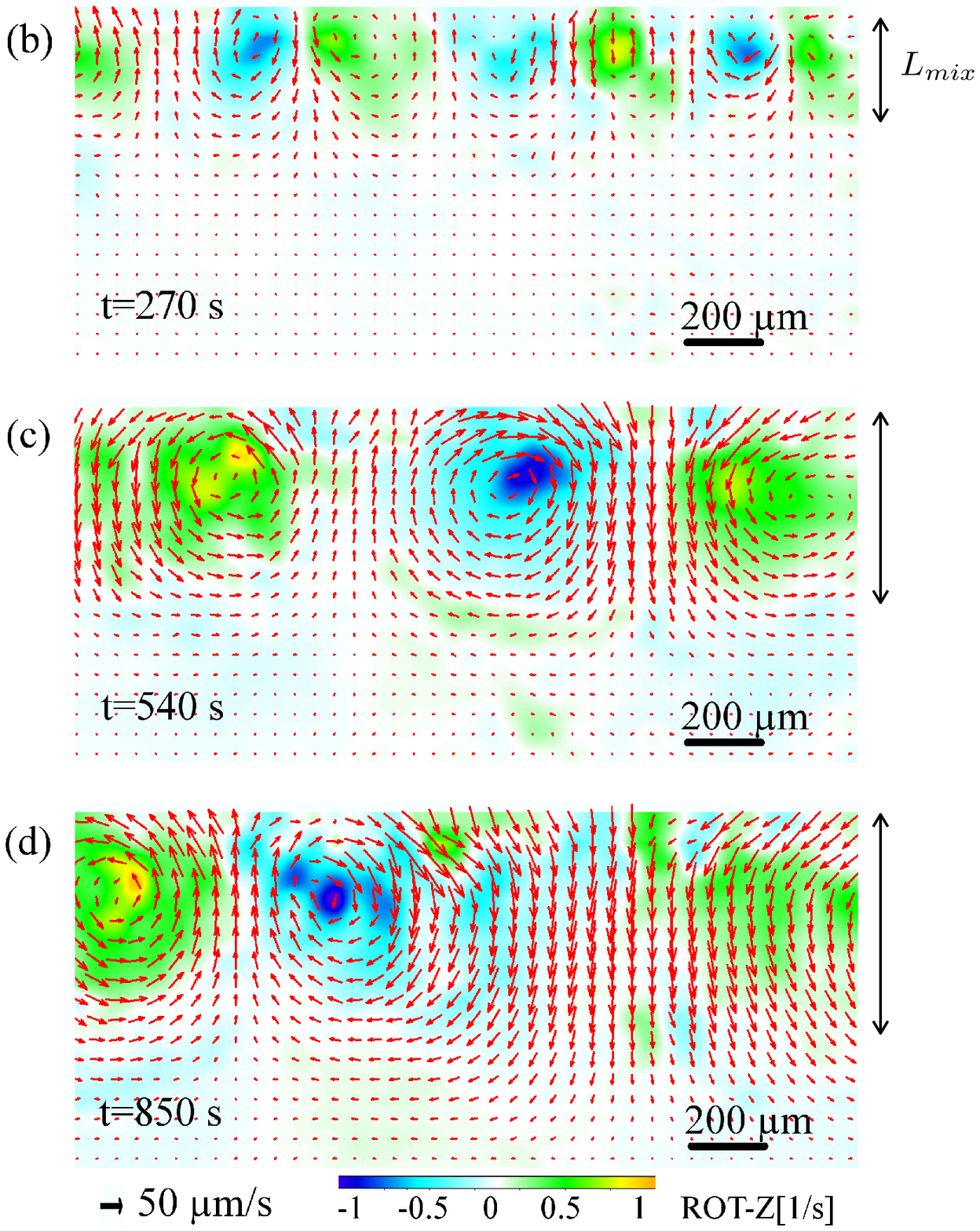}
\caption{(a) Simultaneous measurements of the dynamics of vertical micro-vortex size (\textcolor{blue}{$\bullet$}) and voltage response $\Delta V$ (\textcolor{red}{{\bf--}}), under a constant current density of 10~A/m$^{2}$ across the charge selective membrane. Four characteristic transport regimes are delineated by the dashed lines. \textrm{I}. electric Ohmic conduction without hydrodynamic convection,  \textrm{II}. a potential jump with the development of ICP,  \textrm{III}. a hydrodynamic convection regime with linearly growing electric resistance and vortex size in time, and IV. a saturated regime with saturated values of both vortex size and electric resistance. 
(b--d) The corresponding flow field, velocity vectors and vorticity ($\nabla \times \vec{u}$) obtained with a PIV technique at different times; (b) and (c) show the growth and (d) the unsteady nature of the micro-vortices. The vertical arrow in (b--d) indicates the length scale of vortex mixing layer, $L_{mix}$, which initially increases with time. $L_{mix}$ is measured to be 260, 480, and 550~$\mu m$ from (b) to (d), respectively, at different times ($t$) indicated.} 
\label{fig3}
\vspace{-0.2in}
\end{centering}
\end{figure}

The first two characteristic electric responses, $\Delta V_{ohm}$ and the onset of $\Delta V_{c}$, are well understood, however the later-time $\Delta V_{olc}$ and the transport mechanisms causing over-limiting conductance, beyond $\Delta V_{c}$, have been extensively debated~\cite{Nikonenko_2014}. This phenomena is manifested in the additional voltage drop $\Delta V_{olc}$ observed under a constant $I$ or the increasing currents in the conventional current-voltage curves under DC voltages by other studies~\cite{Balster_2007,Krol_1999}. Several mechanisms have been proposed, including water dissociation, hydrodynamic convection, and charge-induced membrane discharge~\cite{Dukhin_1991, Rubinstein_2005, Nikonenko_2010, Dydek_2011, Andersen_2012, Nikonenko_2014, Rubinstein_2015}.  Previous work has shown that water dissociation for our type of CMX membrane is insufficient to account for the observed over-limiting conductance~\cite{Balster_2007,Krol_1999,Nikonenko_2014}. To gain insight, we analyze the coupled hydrodynamics from the captured images, using particle pathlines (Fig.~\ref{fig1} b--d) and PIV analysis (Fig.~\ref{fig3} b--d).

In Fig.~\ref{fig3}, a representative set of the coupled dynamics of electric response $\Delta V(t)$ and vertical vortex size $L_{mix}$ (measured from the membrane surface) is shown. The error bars represent the variation of the individual vortex sizes.
Four distinct regimes are delineated in Fig.~\ref{fig3}. The initial Regime \textrm{I} is electric Ohmic conductive, with no hydrodynamic convection observed from the motion of the micro-particles. In Regime \textrm{II}, voltage jump $\Delta V_c$ occurs, with an increasing electrical resistance of the electrodialysis cell. This starts at the transition time (e.g., $\tau_{c} \approx 77$ s for $J = 10$~A/m$^{2}$). In Regime \textrm{III}, small counter-rotating vortex pairs appear along the membrane surface. The thickness of this mixing vortex layer $L_{mix}$ grows linearly in time (e.g. Fig.~\ref{fig3}b). Simultaneously, $\Delta V(t)$ gradually grows in this regime. Finally, in Regime \textrm{IV}, both $\Delta V$ and vortex size saturate and fluctuate at fixed values. The vortices are observed to move laterally and merge together, showing unsteady dynamics (see the supporting videos~\cite{Suppl_info}). Consistent with our experimental findings, the unsteady and chaotic movements of saturated vortexes have been observed in recent direct numerical simulations, where instead of a constant current, a constant voltage drop is the control parameter and the current fluctuates around a saturated value~\cite{Druzgalski_2013,Demekhin_2013}.

\begin{figure}
\includegraphics[width=0.8\columnwidth]{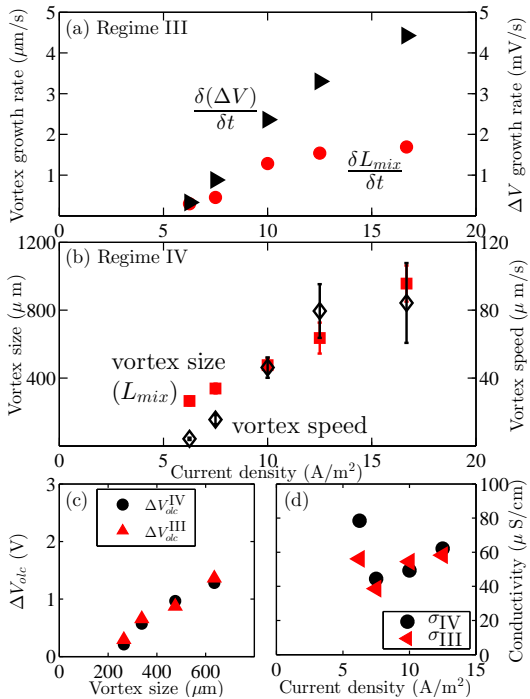} 
\caption{(a) The growth rate of vortex size (\textcolor{red}{$\bullet$}) and voltage drop $\Delta V$ (\textcolor{black}{$\blacktriangleright$}) during the linear convective Regime \textrm{III} indicated in Fig. 3, under different current densities. (b) The dependence of the average vortex size ($L_{mix}$) (\textcolor{red}{$\blacksquare$}) and speed (\textcolor{black}{$\Diamond$}) on the applied current density across the membrane in the saturated convective regime IV. The error bar shows the standard deviation of the time averaged values, revealing more fluctuations in vortex dynamics at higher currents. (c) The average voltage drops, $\Delta V_{olc}$, over the saturated vortex size in Regime \textrm{IV}, ($\bullet$) obtained from the growth rate data of Regime \textrm{III} (\textcolor{red}{$\blacktriangle$}) and from the data of saturated $\Delta V_{olc}^{IV}$ in Regime \textrm{IV} ($\bullet$). (d) The average conductivity in the mixing layer in the growth regime, $\sigma_{\textrm{III}}$ (\textcolor{red}{$\blacktriangleleft$}), and in the saturated regime $\sigma_{\textrm{IV}}$ ($\bullet$).}
\label{fig4} 
\vspace{-0.2in}
\end{figure}

 From our experimental data, micro-vortices only set in from Regime~\textrm{III} , slowly growing in size and speed with time, accompanied by a linear increase of $\Delta V$ in time (e.g. 150 --450 s in~Fig.~\ref{fig3}a). In addition, our experimental result of rms vortex velocity is quantitatively consistent with that found in 2D simulations of electro-osmotic instability under similar electrical forcing (shown in Fig. 4c in Ref.~\cite{Rubinstein_2008}), albeit different electrical boundary conditions (constant current {\it vs.} constant voltage). Based on these observations, the convective transport carried by swirling micro-vortexes is very likely the main cause of the over-limiting conductance observed. However, the fundamental origin of the convective instability, which can be induced via an equilibrium or non-equilibrium mechanism as suggested by different theories~\cite{Dukhin_1991, Rubinstein_2005,Nikonenko_2014, Rubinstein_2015}, remains elusive. Our data revealing quantitative growth of $\Delta V_{olc}$ and vortex speed and size can motivate future theoretical investigations under constant currents to identify the primary origin of the convective instability.

We now show, for the first time, the dependence of electroconvective dynamics on the forced current density in Fig.~\ref{fig4}.  In Regime \textrm{III}, both the voltage $\Delta V_{olc}$ and the mixing layer thickness of vortex region $L_{mix}$ initially grow linearly (see Fig.~\ref{fig3}a). These growth rates versus applied current densities are shown in Figure~\ref{fig4}a. Fig.~\ref{fig4}b displays the average vortex speed and size in the saturated, over-limiting Regime \textrm{IV}, where unsteady dynamics are observed. Both vortex size and speed increase with current density, but the rms velocity increases stronger than the vortex size for the increasing current, underlining the importance of convective transport in this regime. The large error bars at the higher current densities reflect the increasing fluctuations and chaotic motions of the individual vortices.

From the data of growth rates in Fig.~\ref{fig4}a, the voltage difference required to grow to a vortex region $L_{mix}$ can be calculated using Regime \textrm{III} data: $\Delta V_{olc}^{III}=(\delta(\Delta V)/\delta t)(\delta L_{mix}/\delta t)^{-1}L_{mix}$. Revealed in Fig.~\ref{fig4}c, this voltage difference estimated for the final saturated vortex region agrees well with the voltage drop $\Delta V_{olc}$ over the saturated mixing layers observed in Regime \textrm{IV}: $\Delta V_{olc}^{IV} = \Delta V - IR_{ohm} - \Delta V_{c}$, indicating that micro-vortices only set in with an excess voltage drop, $\Delta V_{olc}$, across the mixing layer of vortex region. Furthermore, one could estimate the electrical conductivity $\sigma$ in this mixing region due to the presence of swirling vortices: $\sigma_{\textrm{IV}}=L_{mix} J /\Delta V_{olc}^{IV}$, for each $J$. This conductivity is approximately constant for different current densities, and similar to the conductivity of the mixing layer in the growth regime \textrm{III}:  $\sigma_{\textrm{III}} = J (\delta L_{mix}/\delta t)(\delta(\Delta V)/\delta t)^{-1}$, revealed in Fig.~\ref{fig4}d. We found that the conductivity in the mixing layer is similar for all the experiments: $\sigma=55\pm12\,\mu$S/cm. In contrast, in the under-limiting and limiting regimes, without vortices, ion concentrations are slowly depleted due to electrical migration and concentration diffusion, as well as the charge selectivity of the membrane. In the over-limiting regime, micro-vortices are initiated, continue to grow, and finally are saturated, right below the membrane. The viscous dissipation in the mixing layer, $P_{visc}$, was estimated with the gradients of planar velocity field, using a nearest neighbor approach: $P_{visc}=\mu\int (\partial_i v_j )^2~dV$, with the Einstein notation ($i, j =1,2$), volume element $dV$, and $\mu$ the liquid dynamic viscosity  ($1$~mPa$\cdot$s). We assume no shear in the $z$ direction to estimate the volume integral of viscous dissipation. Comparing to the electrical power input $P_{in} = P_{elec}= J A_{mem} \Delta V_{olc}$, the ratio of $P_{visc}/P_{in}$ in Regime~\textrm{IV} is on the order of magnitude of $10^{-8}-10^{-9}$. This indicates extremely low power of energy dissipation by convective vortices, and thus $\Delta V_{olc}$ represents the electrical resistance of the mixing layer of low ion concentration. The advection flow modifies the anion concentration gradient by bringing anions towards the membrane, which impairs the early-time ICP (in Regime~\textrm{II}) and sustains over-limiting conductivity. 

In summary, the electrodialysis system involving a charge selective membrane for charge separation in general presents four distinct dynamic regimes observed in chronopotentiometric measurements: (\textrm{I}) a linear Ohmic electric response as charge diffusion and migration takes place; (\textrm{II}) a jump in electrical response ($\Delta V$) during the development of ICP due to charge-selectivity of the membrane; (\textrm{III}) a linear growth regime where micro-vortices grow in both size and speed with time; and finally (\textrm{IV}) a saturated electro-convective regime having saturated values of vortex speed/size and voltage response $\Delta V$. Our quantitative results of the growth rates and saturated electro-convective responses elucidate that micro-vortices only set in with an excess voltage $\Delta V_{olc}$, have small viscous dissipation, and moreover sustain a nearly constant conductivity in the mixing region.

 \vspace{-0.1in}
\begin{acknowledgments}
\vspace{-0.1in}
We thank M. Wessling, A. Benneker, C. Druzgalski, A. Mani and W. van Baak for the scientific discussions. The research was supported in the cooperation framework of Wetsus, centre of excellence for sustainable water technology (www.wetsus.nl). Wetsus is co-funded by the Dutch Ministry of Economic Affairs and Ministry of Infrastructure and Environment, the European Union Regional Development Fund, the Province of Frysl$\hat{a}$n, the Northern Netherlands Provinces and University Campus Frysl$\hat{a}$n. The authors like to thank the participants of the research theme Biomimetic membranes for the discussions and financial support. R.G.H.L. acknowledges the European Research Council for the ERC starting grant 307342-TRAM. P.A.T. acknowledges Natural Sciences and Engineering Research Council of Canada (NSERC) for the Discovery and Accelerator grants (NSERC RGPIN06297 and RGPAS 477919).
\end{acknowledgments}
\vspace{-0.1in}
Email addresses: $^{\S}$ joeri.devalenca@wetsus.nl;~$^{\ast}$ Corresponding author:~peichun.amy.tsai@ualberta.ca

\bibliography{References_v3}

\end{document}